\documentclass[aps,prl,twocolumn,groupedaddress]{revtex4}
\usepackage{graphicx}

\begin{document}

%\preprint{}

\title{Electronic-enthalpy functional for finite systems under pressure}


%\thanks{}

\author{Matteo Cococcioni,$^{1}$ Francesco Mauri,$^{2}$ 
Gerbrand Ceder,$^{1}$ and Nicola Marzari$^{1}$}

\affiliation{$^{1}$Department of Materials Science and Engineering, and 
Institute for Soldier Nanotechnologies, \\
Massachusetts Institute of Technology, 
77 Massachusetts Avenue, Cambridge, MA 02139, USA \\
$^{2}$Laboratoire de Min\'eralogie-Cristallographie de Paris, \\
Universit\'e Pierre et Marie Curie, 
4 Place Jussieu, 75252, Paris, Cedex 05, France}


\date{\today}

\begin{abstract}
We introduce the notion of electronic enthalpy
for first-principles structural and dynamical calculations of
finite systems under pressure. An external pressure field
is allowed to act directly on the electronic structure of the system studied
via the ground-state minimization of the functional $E+PV_{q}$,
where $V_{q}$ is the quantum volume enclosed by a charge isosurface.
The Hellmann-Feynman theorem applies, and assures that the 
ionic equations of motion follow an isoenthalpic dynamics.
%practical implementation in modern electronic-structure codes is 
%straightforward.  
No pressurizing medium is explicitly required, while
coatings of environmental ions or ligands can be introduced if 
chemically relevant.  We apply this novel approach to the study of group-IV
nanoparticles during a shock wave, highlighting the significant differences in
the plastic or elastic response of the diamond cage under load, and their
potential use as novel nanostructured impact-absorbing materials.
\end{abstract}

% insert suggested PACS numbers in braces on next line \pacs{}
%\keywords{}

\maketitle

%\section{}
The study of nanoparticles under pressure is rapidly acquiring
great scientific and technological interest, since it allows to explore 
structural arrangements and phase transitions under different kinetic and
thermodynamic conditions than those of bulk solids.
While experiments on the compression of nanoparticles have been performed
for more than a decade \cite{ali1,ali2,ali3,ali4,ali5},
and encompass broad classes of materials \cite{si3n4,cofe2o4},
computational studies have been hampered by the conceptual
difficulty arising in applying classical or electronic-structure methods
to the study of finite systems under pressure.
Extended systems can be studied using variable-cell
isoenthalpic dynamics \cite{and,pr,rmw,ivo}; however,
these methods do not carry over to the case of finite systems,
unless the environment (i.e. the pressurizing medium) is introduced
explicitly in the calculations \cite{mar1,mar2,curotto}. 
A full quantum-mechanical treatment becomes then too costly in all
but the smallest cases, due to the large number of atoms or molecules that are
needed to reproduce the pressurizing environment; in addition, equilibration of
the system at a given pressure requires extensive simulations averaging over
many collision events.
To solve the first of these problems, 
a mixed quantum-mechanics/molecular-mechanics approach has been
proposed \cite{mar1,mar2} where
the quantum fragment is compressed by a pressurizing medium 
described by classical force fields.

In this Letter we present a novel and general approach to this problem,
and show how pressure can be applied to a finite system without
the need to introduce a pressurizing medium.
We introduce here, as thermodynamic functional describing a finite system
under pressure, the electronic enthalpy 
\begin{equation}
\label{ental}
%H = E [\{\Psi\}]+ PV_{q}[\{\rho\}]{\rm ,}
H = E + PV_{q}{\rm ,}
\end{equation}
where $E$ is the internal energy of the system, $P$ is the desired
pressure and $V_{q}$ is the ``quantum volume'' occupied by the
electronic charge density.
This quantity is well defined once the threshold for a
density isosurface is chosen, and
can be straightforwardly computed as
\begin{equation}
\label{vq}
V_{q} = \int d{\rm \bf r} \hspace{2truemm}
\vartheta(\rho({\rm \bf r})-\alpha)\;,
\end{equation}
where $\rho({\rm \bf r})$ is the electronic density and
$\vartheta$ is a step function 
at the threshold value $\alpha$. For computational convenience
we smooth the $\vartheta$ to $\tilde \vartheta$, defined as the integral of a 
normalized Gaussian of width 
$\sigma = \alpha/3$ \cite{ftn}. 

In density-functional theory, the total potential acting on the electrons will be
given by the functional 
derivative of the electronic enthalpy in (\ref{ental}) with respect
to the charge density $\rho({\rm \bf r})$. This results
in an additional contribution deriving from the term $PV_{q}$ which is simply
\begin{equation}
\Phi_{V}({\rm \bf r}) = P \frac{\delta V_{q}}{\delta \rho}|_{\rho=
\rho({\rm \bf r})} = \frac{P}{\sigma\sqrt{2\pi}}
e^{-(\rho({\rm \bf r})-\alpha)^{2}/2\sigma^{2}}\;.
\end{equation}
In the self-consistent solution of the Kohn-Sham equations 
the $ \Phi_V $ potential drives the evolution of the electronic
ground-state to the minimum of the electronic enthalpy (\ref{ental}).
The Hellmann-Feynman theorem applies, and thus
the ionic forces take implicitly into account the contribution of the
external load, transferring the effects of the compression 
directly to the ionic relaxation and dynamics. A Lagrangian 
formulation (e.g. for first-principles Car-Parrinello molecular dynamics
\cite{cp}) also follows directly.

The introduction of an electronic-enthalpy functional has
several conceptual and practical advantages:
1) The pressure field acts directly on the electrons; thus, the 
compressibility of the system is properly dominated by electrostatic and
Pauli-principle effects. Also,
if pressure were transferred directly to the ionic nuclei
by an external classical force field, electrons could ultimately
be squeezed out of the system.
2) Isoenthalpic relaxations and dynamics do not require extensive
equilibration with a pressurizing medium; a medium can always be introduced
as an environmental coating, if chemically relevant.
3) In the thermodynamic limit, the classical 
macroscopic volume and enthalpy are recovered.
4) Implementation in electronic-structure codes is straightforward, 
and can be easily extended to other wavefunction-based 
approaches such as Hartree-Fock or variational 
quantum Monte Carlo.
(The results presented here have all been obtained using
first-principles Car-Parrinello molecular dynamics \cite{cp}, as 
implemented in the public domain \newline$\nu$-ESPRESSO package \cite{code}.)

Other developments and applications
stemming from this formulation can be envisioned.
A surface tension term $H \rightarrow
H + \gamma S$ 
can be introduced, where the surface area is obtained by finite differences via
\begin{equation}
\label{asurf}
S\,=\,\int\,d{\rm \bf r}\,[\tilde \vartheta(\rho({\rm \bf r})-(\alpha - \frac{\Delta}{2})) -
\tilde \vartheta(\rho({\rm \bf r})-(\alpha + \frac{\Delta}{2}))]
\frac{|\overrightarrow{\nabla} \rho({\rm \bf r})|}{\Delta}
\end{equation}
and $\Delta$ determines the width of the discretization.
This latter contribution to the energy functional 
could be used to characterize surface tension in metallic nanoparticles,
confinement effects in electron 
bubbles (e.g., for electrons in superfluid helium \cite{silvera}) 
or cavitation effects in solvation models \cite{gygi}.
Finally, we stress that although classical definitions of volume can be introduced in atomistic
simulations \cite{berry,calvo}, a natural measure based on the electron density 
can only emerge in a quantum-mechanical formalism. 
\begin{figure}%[!t]
\includegraphics[width=7.5truecm]{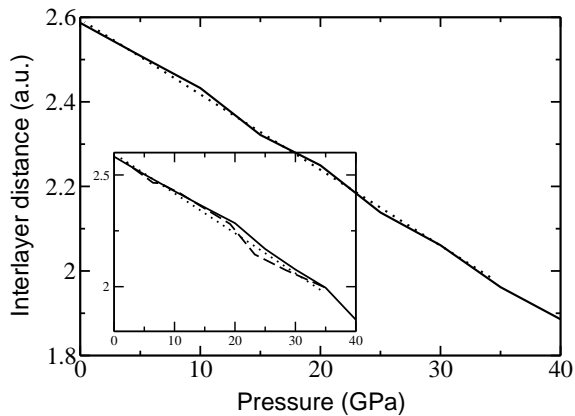}
\caption{\label{dofp}The dependence of the innermost interlayer distance on 
pressure for a 16-layer silicon slab relaxed using the 
electronic-enthalpy functional (solid curve); results from 
the Murnaghan equation of
state for a 16-layer bulk supercell (dotted) are shown for comparison. 
The inset shows the results obtained for a 8-layer slab (solid),
a 8-layer supercell (dotted), and a 8-layer slab compressed via the
explicit introduction of argon as a pressurizing medium (dashed).}
\end{figure}
A validation of 
the present approach can be provided not only in the 
thermodynamical limit (where the quantum volume V$_q$ becomes 
indistinguishable from the classical limit), but also for finite, small systems.
For this purpose we studied the effect of pressure on thin silicon slabs. 
This reference system allows for a direct comparison 
with bulk results without introducing additional
surface tension effects.
We plot in Fig. \ref{dofp} the average distance between the innermost layers in
two Si(001) slabs of 8 and 16 layers, as a function of the
pressure parameter P entering our functional (\ref{ental}).
A remarkable agreement is found with the results obtained when calculating the physical 
pressure \cite{NM} in a bulk silicon system having the same geometry and
compressed along the [001] direction. For the 8-layer slab a full 
quantum-mechanical calculation taking into account explicitly argon
as a pressurizing medium has been performed, delivering an even 
closer agreement (at least up to 20 GPa) 
to our electronic-enthalpy results.
This agreement clearly demonstrates that the minimization of the 
quantum enthalpy correctly accounts for the effects of pressure; 
in addition, agreement is reached rapidly with respect 
to system size \cite{ftn2}.

\begin{figure}%[!t]
\includegraphics[width=7.8truecm]{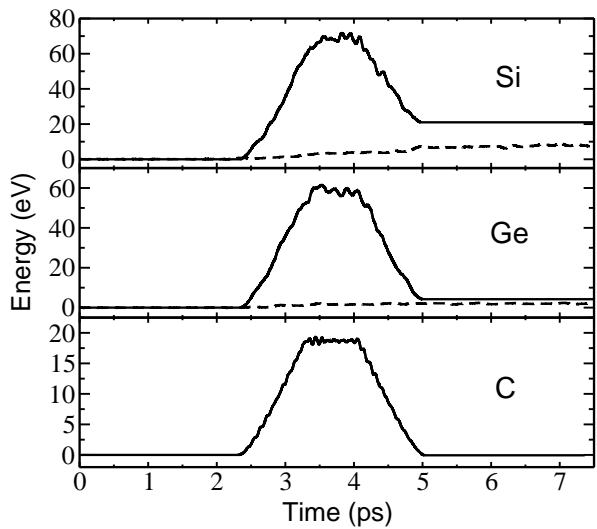}
\caption{\label{work}Work $\int P dV$ done by the external pressure
on the nanoparticles during the shock wave (solid line), and instantaneous
kinetic energy of the ions (dashed line, zero corresponds to the 300 K average).}
\end{figure}

Our case studies are group-IV nanoparticles
M$_{35}$H$_{36}$ during a shock wave (M corresponds to Si, Ge and C).
The initial configurations, carved from bulk diamond, have 
all dangling bonds saturated with hydrogens.
The structures were relaxed and thermalized for 2-3 ps
at 300 K and zero pressure, at which point the ionic thermostat was 
switched off (no electronic thermostat is needed in these simulations).
We modeled the adiabatic
compression characterizing a shock wave with
a linear increase in hydrostatic pressure 
from 0 to 40 GPa for 1 ps, followed by 
0.75 ps at constant pressure and by a linear decrease to 0 GPa in 1 ps.
The system was finally allowed to evolve at zero pressure for another 8-10 ps.
\begin{figure}%[!t]
\includegraphics[width=7.42truecm]{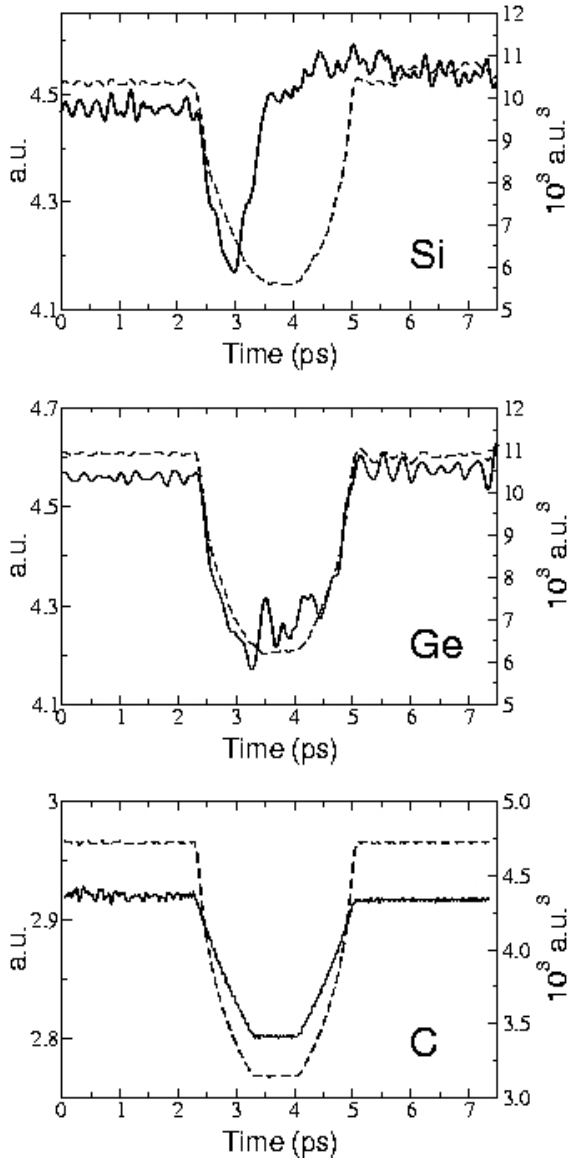}
\caption{\label{blen}Time evolution of the average bond length (solid line) 
and of the quantum volume (dashed line) for the
three nanoparticles during the shock wave.}
\end{figure}
\begin{figure}%[!t]
\includegraphics[width=7.42truecm]{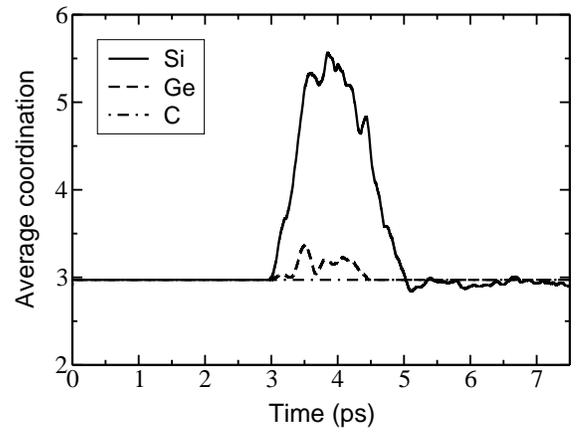}
\caption{\label{coord}Time evolution of the average ionic coordination
for Si (solid line), Ge (dashed line) and C (dot-dashed line)
nanoparticles during the shock wave.}
\end{figure}
The structural transformations occurring during the shock are highlighted
in Figs. \ref{blen} and \ref{coord},
where the time evolution of the average bond length,
quantum volume, and average coordination of the 
ions are shown for the three nanoparticles.
The response of the Si nanoparticle 
(see Fig. \ref{blen})
is remarkable:
a transition is induced 
$\sim$0.6 ps into the compression (P $\simeq$ 26-28 GPa), at which point
the system amorphizes. In this process
the second shell of neighbors overlaps and merges with the first,
increasing the coordination of the ions (see Fig. \ref{coord})
from 3.0 - due to the undercoordinated surface atoms - to $\sim$5.5.
The compressed bonds recover, albeit only on average, the zero-pressure
equilibrium length; volume contraction is achieved
by a collapse of the coordination shells rather than a 
shortening of the nearest bonds.
The resulting amorphous configuration remains stable after pressure is
released, and no recrystallization is detected in the following
8-10 ps.  A structural relaxation confirms that
the amorphous nanoparticle corresponds to a well-defined local minimum 
for the potential energy surface; this minimum is 
11.15 eV (0.32 eV per silicon atom)
higher than the corresponding initial tetrahedral configuration.
These results confirm the existence of an amorphous phase for silicon
nanoparticles, as reported in Ref. \cite{ali5} 
and also found in the simulations of Refs. \cite{mar1,mar2}.
Germanium and carbon at these sizes display instead an
elastic response (see Fig. \ref{blen}) and 
the original tetrahedral structure is recovered at the end of the simulation. 
Due to its very stiff bonds, the carbon nanoparticle maintains its tetrahedral 
coordination even at peak pressure, while germanium is noticeably softer.
In both cases, however, the average coordination of the ions is barely affected
(see Fig. \ref{coord}) and the first two peaks in the pair-correlation function
remain distinct at all times during the shock.
The differences observed in the structural behavior 
are also reflected in the thermodynamics of the compression process.
In Fig. \ref{work} we show the mechanical work exerted on the 
nanoparticles by the compression wave; 
in our approach this can be straightforwardly determined 
by $\int P dV$, whereas the fluctuations of an explicit pressurizing 
medium would hinder the volume changes
of the quantum fragment.
It is immediately evident that the Si nanoparticle absorbs a great amount
of mechanical energy: a portion of this energy is converted into 
kinetic energy of the ions, 
thanks to enhanced anharmonic coupling between 
acoustic and optical phonons, while a larger part goes into 
the plastic deformation and
transition into the metastable amorphous phase.
Both the diamond nanoparticle and 
(to a lesser extent) the germanium one 
release instead elastically all the mechanical energy absorbed during the
shock and stored in the compression of the covalent bonds.

Kinetic trapping for small Si nanoparticles
has been demonstrated in Ref. \cite{ali5},
with an amorphous phase appearing
during pressure release below 5 GPa. In agreement with Ref. \cite{ali5},
we find an enhancement with respect to the bulk for the
pressure transition during the upstroke phase; our nanoparticle falls
directly in the amorphous phase, without an intermediate
$Imma$ or $\beta$-Sn phase.
Pressure-induced amorphization was observed 
when studying porous silicon films containing diamond-structure 
nanocrystals under load \cite{natl};
amorphization was also found in bulk first-principles simulations 
\cite{duran,morishita}
in the 11-15 GPa range.
The importance of the rate of load release
on the structural transformation in Si and Ge 
was highlighted in Ref. \cite{mujica}. 
We have verified that changing the speed of the shock wave 
by a factor of 1/15, 1/2, or
the initial temperature of the system from 300 K to 100 K,
did not change the amorphization pathway or the transition pressure. 
We also carefully confirmed that the nanoparticle never becomes liquid.

Finally, we examined the effects that the threshold parameter $\alpha$ 
defining the quantum volume has on the isoenthalpic dynamics. 
The silicon simulation was repeated using 
$\alpha = 0.0005$ electrons/(a.u.)$^{3}$
($\alpha = 0.0002$ electrons/(a.u.)$^{3}$ was 
used for all previous simulations).
While at 0 GPa this different 
choice leads to a volume difference of $\sim 10\%$,
in the course of the compression the two definitions rapidly 
converge to the same value;
the soft ``electronic skin'' involved when
using a lower threshold 
contracts quickly, and 
the system then opposes the same resistance to the external load
as in the case of a higher threshold.
More importantly, the energy absorbed is shown to be the same independently of 
our choice of threshold; the outer ``electronic skin''
releases elastically at the end of 
the run any energy stored at the beginning of the
simulation.

In conclusion, we have introduced a novel approach in the study of 
finite systems under pressure, based on the quantum-mechanical definition of an
electronic-enthalpy functional. This method is straightforward to implement in static
and dynamical first-principles calculations, and opens the study of the
phase stability and kinetics of finite quantum systems to extensive first-principles
simulations.
Its accuracy and broad independence on the isosurface threshold have also been
established.  We have studied
the behavior of semiconducting nanoparticles (M$_{35}$H$_{36}$, with M = Si, Ge, and C)
during a hydrostatic shock wave, highlighting the different
response of the three materials. Small
silicon nanoparticles undergo plastic amorphization at pressures
larger than those required to amorphize a bulk system, while carbon nanoparticles
(and to a lesser extent germanium ones)
elastically recover their original tetrahedral
structure \cite{ftn3}. 
A large amount of mechanical energy is absorbed
by silicon nanoparticles as internal energy of their amorphous phase and 
kinetic energy of the ions. This energy will be released in longer timescales than those of a 
traveling shock wave, hinting at the possibility of designing a nanostructured
impact-absorbing material.
It is also
envisioned that the transition pressures, decay timescales, and the overall mechanical
response could be tuned by an appropriate choice of materials, composition, and sizes.
Support from MIT Institute for Soldier Nanotechnologies, grant DAAD19-02-D-0002, is gratefully acknowledged.


\begin{thebibliography}{999}

\bibitem{ali1} S. H. Tolbert and A. P. Alivisatos, Science {\bf 265}, 373
(1994).
                                                                                
\bibitem{ali2} S. H. Tolbert and A. P. Alivisatos, J. Chem. Phys. {\bf 102},
4642 (1995).
                                                                                
\bibitem{ali3} C. Chen, A. B. Herhold, C. S. Johnson, A. P. Alivisatos, Science
{\bf 276}, 398 (1997).
                                                                                
\bibitem{ali4} J. N. Wickham, A. B. Herhold, A. P. Alivisatos, Phys. Rev. Lett.
{\bf 84}. 923 (2000).
                                                                                
\bibitem{ali5} S. H. Tolbert, A. B. Herhold, L. E. Brus, A. P. Alivisatos,
Phys. Rev. Lett. {\bf 76}, 4384 (1996).
                                                                                
\bibitem{si3n4} Z. Wang et al., Appl. Phys. Lett. {\bf 83}, 3174 (2003).
                                                                                
\bibitem{cofe2o4} Z. Wu et al., J. Appl. Phys. {\bf 93}, 9983
(2003).

\bibitem{and} H. C Andersen, J. Chem. Phys. {\bf 72}, 2384 (1980).

\bibitem{pr} M. Parrinello and A. Rahman, Phys. Rev. Lett. {\bf 45}, 
1196 (1980); J. Appl. Phys. {\bf 52}, 7182 (1981).

\bibitem{rmw} R. M. Wentzcovitch, Phys. Rev. B {\bf 44}, 2358 (1991).

\bibitem{ivo} I. Souza and J. L. Martins, Phys. Rev. B {\bf 55}, 8733 (1997).

\bibitem{mar1} R. Marton\'ak, C. Molteni, M. Parrinello, Phys. Rev. Lett.
{\bf 84}, 682 (2000); C. Molteni, R. Marton\'ak, M. Parrinello, J. 
Chem. Phys. {\bf 114}, 5358 (2001).

\bibitem{mar2} R. Marton\'ak, C. Molteni, M. Parrinello, J. 
Chem. Phys. {\bf 117}, 11329 (2002). 

\bibitem{curotto} E. Curotto, J. Chem. Phys. {\bf 114}, 10702 (2001).

\bibitem{ftn} The smearing width is largely irrelevant in determining the physical
response of the system.

\bibitem{cp} R. Car and M. Parrinello, Phys. Rev. Lett. {\bf 55}, 2471 (1985).

\bibitem{code} S. Baroni, A. Dal Corso, S. de Gironcoli, P. Giannozzi, C. Cavazzoni, G. Ballabio, S. Scandolo, G. Chiarotti, P. Focher, A. Pasquarello, K. Laasonen, A. Trave, R. Car, N. Marzari, A. Kokalj, http://www.pwscf.org/.

\bibitem{silvera} J. Tempere, I. F. Silvera, J. T. Devreese, Phys. Rev. Lett.
{\bf 87}, 275301 (2001); M. Rosenblit and J. Jortner, Phys. Rev. Lett.
{\bf 75}, 4079 (1995).

\bibitem{gygi} J. L. Fattebert, F. Gygi, J. Comput. Chem. {\bf 23},
662 (2002); D. Scherlis {\it et al.} (to be published).

\bibitem{berry} H.-P. Cheng, X. Li, R. L. Whetten, and R. S. Berry, Phys. 
Rev. A {\bf 46}, 791 (1992).

\bibitem{calvo} F. Calvo and J. P. K. Doye, Phys. Rev. B {\bf 69}, 125414 
(2004)

\bibitem{NM} O. H. Nielsen and R. M. Martin, Phys. Rev. Lett. {\bf 50}, 697
(1983); Phys. Rev. B {\bf 32}, 3780 (1985); Phys. Rev. B {\bf 32}, 3792 (1985).

\bibitem{ftn2} We stress that no fitting procedure has been used to find the
agreement among the two curves; the pressure parameter on the graph was
the one used in the electronic-enthalpy functional.

\bibitem{natl} S. K. Deb et al., Nature {\bf 414}, 528 (2001).

\bibitem{duran} M. Durandurdu and D. A. Drabold, Phys. Rev. B {\bf 67},
212101 (2003).

\bibitem{morishita} T. Morishita, Phys. Rev. Lett. {\bf 93}, 55503 (2004).

\bibitem{mujica} A. Mujica et al., Rev. Mod. Phys. {\bf 75}, 863 (2003).

\bibitem{ftn3} Larger Ge nanoparticles can also undergo 
some amorphization. A detailed description of these size effects will be 
published separately.

\end{thebibliography}
\end{document}